\documentclass[english,aps,prb,twocolumn,showpacs]{revtex4-1}
\usepackage[T1]{fontenc}
\usepackage[latin9]{inputenc}
\usepackage{amsmath}
\usepackage{amssymb}
\usepackage{graphicx}
\usepackage{bm}
\usepackage{babel}
\usepackage{color}

\makeatletter

 \@ifundefined{textcolor}{}
 {
   \definecolor{BLACK}{gray}{0}
   \definecolor{WHITE}{gray}{1}
   \definecolor{RED}{rgb}{1,0,0}
   \definecolor{GREEN}{rgb}{0,1,0}
   \definecolor{BLUE}{rgb}{0,0,1}
   \definecolor{CYAN}{cmyk}{1,0,0,0}
   \definecolor{MAGENTA}{cmyk}{0,1,0,0}
   \definecolor{YELLOW}{cmyk}{0,0,1,0}
 }

\makeatother

\begin{document}

\title{Transport anomalies due to anisotropic interband scattering}

\author{Maxim Breitkreiz}
\affiliation{Institute of Theoretical Physics, Technische Universit\"at Dresden,
01062 Dresden, Germany}

\author{P. M. R. Brydon}
\affiliation{Institute of Theoretical Physics, Technische Universit\"at Dresden,
01062 Dresden, Germany}

\author{Carsten Timm}
\email{carsten.timm@tu-dresden.de}
\affiliation{Institute of Theoretical Physics, Technische Universit\"at Dresden,
01062 Dresden, Germany}

\date{July 12, 2013}

\begin{abstract}
Unexpected transport behavior can arise due to anisotropic
single-particle scattering in multiband systems. Specifically, we show within a
semiclassical Boltzmann approach beyond the relaxation-time approximation that
anisotropic scattering between electronlike and holelike Fermi surfaces
generically leads to negative transport times, which in turn cause
negative magnetoresistance, an extremum in the Hall coefficient, and a
reduction of the resistivity. The anisotropy required for this to occur
decreases with increasing mismatch between the Fermi-surface radii.
\end{abstract}

\pacs{
72.10.-d,
72.15.Lh,
74.70.Xa
}

\maketitle

\section{Introduction}

In basic transport theory, the acceleration
of carriers by an applied electric field is balanced by scattering,
leading to a steady state with constant drift current.\cite{Ziman1972}
The direction of the drift current is determined by the charge of the
carriers and relaxes over a characteristic transport time (TT).
Deviations from this basic picture tend to generate a lot of interest.
A striking example are \emph{negative} TTs: Some fraction of the
carriers may drift in the direction opposite of what one would expect based on
their charge.
Negative TTs of minority carriers were predicted for systems with
electron- and holelike Fermi surfaces (FSs),\cite{key-2} based on strong
electron-hole (two-particle) scattering. This \emph{carrier drag} was
first observed in semiconductor quantum wells.\cite{oldcarrierdrag}

In this paper, we explore a different origin of negative TTs and its
consequences for transport: Negative TTs can also arise due to anisotropic
single-particle scattering\cite{key-3} in multiband systems.
In the case of \emph{isotropic} scattering, the TTs are equivalent
to the usual quasiparticle lifetimes. Anisotropic scattering, on the
other hand, favors certain scattering angles.
As a result, the transport coefficients can deviate significantly from the
expectation based on the bare lifetimes.

Anisotropic single-particle scattering in multiband systems
can be realized in materials close to an
excitonic instability, e.g., electron-hole bilayers,\cite{Hu2000} the
iron pnictides,\cite{CEE08,HCW08,BrT09} 
chromium and its alloys,\cite{HaR68,Cralloy} and the transition-metal
dichalcogenides.\cite{LHQ07,CMC07} In these materials,
nesting of electron and hole FSs strongly enhances interband spin or
charge fluctuations with wavevector close to the nesting vector ${\bf Q}$.
These fluctuations are expected to promote highly anisotropic scattering
between the nested FSs.

In the iron pnictides, the effect of such scattering seems to be especially
pronounced, as the normal-state transport coefficients show highly
anomalous behavior. In particular, the unexpectedly small
magnetoresistance is hard to reconcile with the strongly enhanced Hall
coefficient if analyzed based on
a simple multiband model with \emph{positive} transport times.\cite{Kim}
In these materials, anisotropic scattering is thought to be mainly due to
spin fluctuations.\cite{Prelovsek,qplifetime,OnK12,Fanfarillo2012}
Fanfarillo \textit{et al.}\ \cite{Fanfarillo2012} have demonstrated that
vertex corrections can lead to an enhancement of the Hall coefficient,
which could explain its pronounced temperature dependence in the
pnictides.\cite{Fang2009,Shen2011}
Vertex corrections result from the anisotropy of the scattering and
are responsible for the difference between bare lifetimes and TTs.
They also show that the enhancement of the Hall coefficient
is connected to negative TTs of minority carriers.\cite{Fanfarillo2012}

In the present work we focus on the effect of anisotropic interband
scattering on transport coefficients.
Working within a Boltzmann approach beyond the relaxation-time approximation,
we consider two FSs and include an interband scattering rate
with arbitrary dependence on the scattering angle. We show that if
one FS is electronlike and the other holelike, anisotropy not only
leads to an enhancement of the Hall coefficient\cite{Fanfarillo2012}
but also to an extremum. Moreover, we find a reduction of the resistivity
and negative magnetoresistance as direct
consequences of negative TTs.

\section{Model and method}

\begin{figure}[t]
\includegraphics[width=\columnwidth]{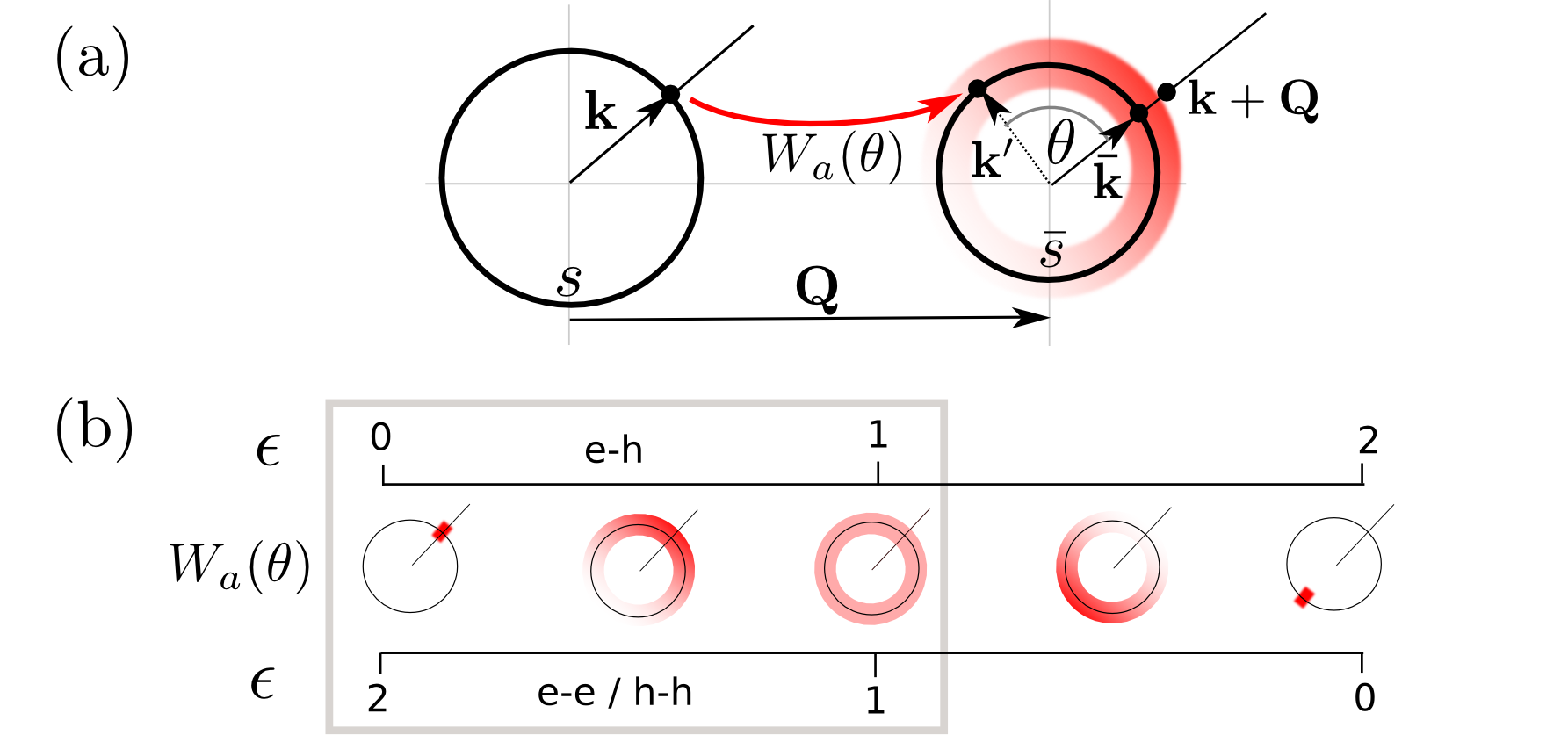}
\caption{(Color online) (a) Sketch of the two isotropic FSs $s$ and
$\bar{s}$, displaced by $\mathbf{Q}$, and the elastic
interband scattering rate $W_{a}(\theta)$ (red/gray gradient).
(b) Illustration of the relation between the anisotropy parameter
$\epsilon$ and the shape of the function $W_{a}(\theta)$, for the case of
$W_a(\theta)$ having a single peak. We focus on the situation 
where the peak appears at $\theta=0$, highlighted by the box. }
\label{fig:formulation}
\end{figure}

We consider
2D and 3D metals with two isotropic FSs labeled by $s=1,2$. The center
of one FS is displaced with respect to the other by a wave vector
$\mathbf{Q}$, see Fig.~\ref{fig:formulation}(a). One of the FSs may be electronlike and the other
holelike (e-h case) or both may be of the same type (e-e/h-h case).
We employ a semiclassical description in terms of the distribution
function $f_{s,\mathbf{k}}=f_{0}(\varepsilon_{s,\mathbf{k}})+g_{s,\mathbf{k}}$,
with the equilibrium Fermi-Dirac distribution
$f_{0}(\varepsilon_{s,\mathbf{k}})$, where $\varepsilon_{s,\mathbf{k}}$ is the band energy,
and a deviation $g_{s,\mathbf{k}}$.
In a weak uniform electric field $\mathbf{E}$, the stationary state is described
by the linearized Boltzmann equation \cite{Ziman1972}
\begin{equation}
e\mathbf{E} \cdot \mathbf{v}_{s,\mathbf{k}}
  \left[-f_{0}'(\varepsilon_{s,\mathbf{k}})\right]
  = \sum_{s',\mathbf{k}'}
  W_{s,\mathbf{k}}^{s',\mathbf{k}'}\, (g_{s,\mathbf{k}}
  - g_{s',\mathbf{k}'}) ,
\label{eq:BE}
\end{equation}
where $\mathbf{v}_{s,\mathbf{k}} = \hbar^{-1}\,\nabla_{\mathbf{k}}\varepsilon_{s,\mathbf{k}}$ is the velocity,
and $W_{s,\mathbf{k}}^{s',\mathbf{k}'}$ is the scattering rate from state
$s,\mathbf{k}$ to state $s',\mathbf{k}'$. 
The scattering term contains an in-scattering contribution proportional to
$g_{s',\mathbf{k}'}$, which is equivalent to vertex corrections.

Specifically, we consider elastic scattering with an isotropic
intraband contribution $W_{i}$ and an in general anisotropic interband
contribution $W_{a}(\theta_{\mathbf{k},\mathbf{k'}})$,
\begin{equation}
W_{s,\mathbf{k}}^{s',\mathbf{k}'} =
  \delta(\varepsilon_{s',\mathbf{k}'}-\varepsilon_{s,\mathbf{k}})\,
  \left[\delta_{s'\bar{s}}\, W_{a}(\theta_{\mathbf{k},\mathbf{k'}})
  + \delta_{s's}\, W_{i}\right] ,
\label{eq:scatrate}
\end{equation}
where $\bar{s}=2$ ($1$) for $s=1$ ($2$).
The interband scattering rate $W_{a}$ in general depends on 
$\mathbf{k}$ and $\mathbf{k}'$.
Since we consider a weak electric field $\mathbf{E}$, we assume the
displacements of the Fermi seas to be small compared to the Fermi momenta
$k_{F,s}$. Then the range of relevant absolute values $|\mathbf{k}|$,
$|\mathbf{k'}|$ is small compared to the range of relevant polar angles, and we
can ignore the dependence of $W_a$ on the former.
By symmetry, $W_a$ can then only depend on the angle 
$\theta\equiv \theta_{\mathbf{k},\mathbf{k'}}$ spanned by
$\mathbf{k}$ and $\mathbf{k}'$, see Fig.\ \ref{fig:formulation}(a),
and is an even function of $\theta$.

The deviations $g_{s,\mathbf{k}}$ solving the Boltzmann equation (\ref{eq:BE})
are linear in $\mathbf{E}$ and can thus be written as
$g_{s,\mathbf{k}} = e\mathbf{E}\cdot\mbox{\boldmath$\Lambda$}_{s,\mathbf{k}}
[-f_{0}'(\varepsilon_{s,\mathbf{k}})]$.
Since the scattering rate $W_{s,\mathbf{k}}^{s',\mathbf{k'}}$
is an even function of $\theta$,
the scattering does not break rotational symmetry and does not
introduce a preferred direction of rotation. Therefore,
$\mbox{\boldmath$\Lambda$}_{s,\mathbf{k}}$ must be parallel to the only vector
appearing in the equation, namely the velocity
$\mathbf{v}_{s,\mathbf{k}}$. Due to rotational symmetry, the prefactor cannot
depend on $\mathbf{k}$. Thus we can write
\begin{equation}
g_{s,\mathbf{k}} = \tau_{s}\,
  e\mathbf{E} \cdot \mathbf{v}_{s,\mathbf{k}}
  \left[-f_{0}'(\varepsilon_{s,\mathbf{k}})\right] ,
\label{eq:relaxtimeansatz}
\end{equation}
where $\tau_{s}$ is the TT for FS $s$, to be determined below.

Inserting Eq.\ (\ref{eq:scatrate}) and performing the energy integration
we obtain for the right-hand side of Eq. (\ref{eq:BE})
\begin{eqnarray}
\sum_{s'} N_{s'} \Big\langle
  [\delta_{s'\bar{s}} W_{a}(\theta) + \delta_{s's}W_{i}]  
\big( g_{s,\mathbf{k}}  - g_{s',\mathbf{k}'}\big)
  \Big\rangle_\theta \, ,
\end{eqnarray}
where $N_s$ is the density of states for FS $s$ and the average over $\theta$
is denoted by
$\left\langle\ldots\right\rangle_{\theta} = \frac{1}{\pi} \int d\theta
\ldots$ for 2D and $\left\langle\ldots\right\rangle_{\theta} = \frac{1}{2}
\int d\theta \sin\theta\ldots$ for 3D. Equation (\ref{eq:relaxtimeansatz})
shows that $g_{s',\mathbf{k}'}$ depends on $\theta$ through the velocity,
which for isotropic FSs can be  written as
\begin{equation}
\mathbf{v}_{s',\mathbf{k}'}=\eta_{s,s'}v_{F,s'}
  \left(\frac{\mathbf{v}_{s,\mathbf{k}}}{v_{F,s}}\,\cos\theta
  + \mathbf{e}_{\bot} \sin\theta\right) ,
\end{equation}
where $v_{F,s}>0$ is the Fermi velocity for FS $s$, $\eta_{ss'} = +1$
($\eta_{ss'} = -1$) if the FSs $s$ and $s'$ are of the same (different) type,
and $\mathbf{e}_{\bot}$ denotes the unit vector perpendicular to ${\bf
    v}_{s,{\bf k}}$ in the plane spanned by ${\bf
    v}_{s,{\bf k}}$ and ${\bf
    v}_{s',{\bf k}'}$.
Since $W_{a}(\theta)$ is an even function of $\theta$, the term proportional to
$\sin\theta$ averages to zero. Using the definition
  Eq.~(\ref{eq:relaxtimeansatz}) and the assumption of elastic scattering,
the term proportional to $\cos\theta$ can  
be written in terms of $g_{s,\mathbf{k}}$.
Factoring out $g_{s,\mathbf{k}}$, we obtain
\begin{eqnarray}
\lefteqn{ e \mathbf{E} \cdot \mathbf{v}_{s,\mathbf{k}}
  \left[-f_{0}'(\varepsilon_{s,\mathbf{k}})\right]
  = g_{s,\mathbf{k}} \sum_{s'} N_{s'} } \nonumber \\
&& {}\times  \Big\langle
  [\delta_{s'\bar{s}} W_{a}(\theta) + \delta_{s's}W_{i}]
  \Big( 1 - \eta_{ss'}\,
  \frac{\tau_{s'} v_{F,s'}}{\tau_s v_{F,s}}\, \cos\theta\Big)
  \Big\rangle_\theta .\quad
\end{eqnarray}
Together with Eq.\ (\ref{eq:relaxtimeansatz}), this implies
\begin{equation}
\frac{1}{\tau_s} = N_s W_i
  + N_{\bar{s}} \left\langle W_a(\theta)\right\rangle_\theta
   - \eta_{s\bar{s}} N_{\bar{s}}\, \frac{\tau_{\bar{s}}
  v_{F,\bar{s}}}{\tau_s v_{F,s}}\,
  \left\langle W_a(\theta)
     \cos\theta \right\rangle_{\!\theta} .
\label{eq:invtaus.general}
\end{equation}
It is useful to define the \emph{anisotropy parameter}
\begin{equation}
\epsilon \equiv
  1 + \eta_{12} \frac{\left\langle
  W_{a}(\theta)\cos\theta\right\rangle_{\theta}}
  {\left\langle W_{a}(\theta)\right\rangle_{\theta}}
\label{eq:eps}
\end{equation}
as a measure of the anisotropy of $W_a(\theta)$.
The limit $\epsilon=0$ ($\epsilon=2$) corresponds
to $W_a(\theta)$ having a $\delta$-function peak at $\theta=0$ ($\theta=\pi$)
for the e-h case, and vice versa for the e-e/h-h case.
$\epsilon=1$ always corresponds to $\langle W_a(\theta)\cos\theta\rangle_\theta
= 0$, which includes the case of isotropic interband scattering. 
Figure \ref{fig:formulation}(b) illustrates the connection between $\epsilon$
and the anisotropic scattering rate for the case of a single maximum.
We focus on the situation where the maximum is at $\theta=0$.
Although our results hold for a general even function
$W_a(\theta)$, for the
sake of clarity we confine the discussion to an anisotropic scattering rate
with a single
maximum at $\theta=0$, see Fig.\ \ref{fig:formulation}(b). Such a maximum
occurs naturally
for the e-h case close to an excitonic instability, due to the scattering by
the enhanced spin or charge fluctuations, allowing $\epsilon\in[0,1]$ to be
tuned by doping or temperature. In particular, we expect that
  $\epsilon\rightarrow0$ as one approaches the excitonic instability. 
Although there is no excitonic instability for the e-e/h-h case,
$\epsilon\in[1,2]$, collective fluctuations can still be enhanced due to the
proximity of nesting.

\section{Transport times}

We first consider pure interband scattering.
Solving Eq.~(\ref{eq:invtaus.general}) for the TTs, we obtain
\begin{equation}
\tau_{s} = \tau_{0,s}\,
  \frac{1+\frac{1-\epsilon}{\epsilon}\,\big(1-\frac{1}{\gamma_{s}}\big)}
  {2-\epsilon} ,
\label{eq:trlifetime}
\end{equation}
where $\tau_{0,s} = N_{\bar{s}}^{-1}\left\langle
W_a(\theta)\right\rangle_\theta^{-1}$ is the bare lifetime for FS $s$, and
$\gamma_{s} \equiv v_{F,s}\tau_{0,s}/v_{F,\bar{s}}\tau_{0,\bar{s}}
= (k_{F,s}/k_{F,\bar{s}})^{d-1}$, where $d$ is the dimension of the system.
Note that only the surface areas of the FSs matter and not their densities of
states.

\begin{figure}[tb]
\includegraphics[width=\columnwidth]{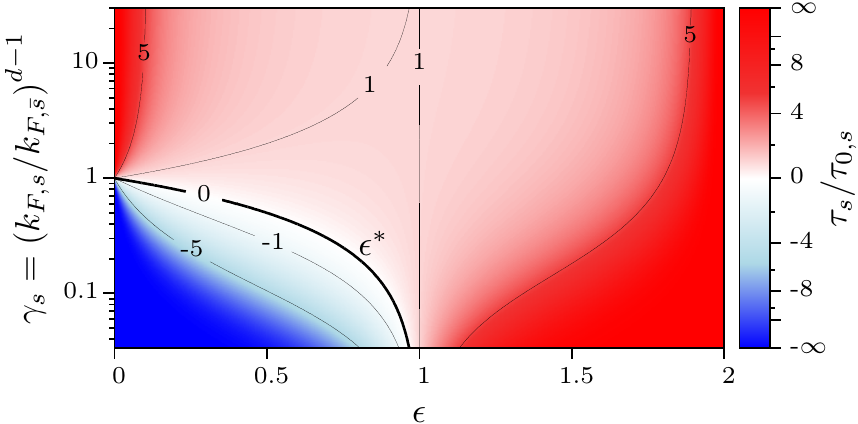}
\caption{(Color online) Transport time $\tau_s$ in units of the bare lifetime
$\tau_{0,s}$ at FS $s$ as a function of the anisotropy parameter $\epsilon$ and
$\gamma_{s}=\left(k_{F,s}/k_{F,\bar{s}}\right)^{d-1}$.
The corresponding plot of $\tau_{\bar{s}}/\tau_{0,\bar{s}}$ would be
the mirror image with respect to $\gamma_{s}=1$.}
\label{fig:tau}
\end{figure}

The TT is plotted in Fig.\ \ref{fig:tau}. We
first focus on the e-h case, $0\le\epsilon\le1$, cf.\ Fig.\
\ref{fig:formulation}(b). The smaller FS has a negative TT for
\begin{equation}
\epsilon < \epsilon^{*} \equiv 1-\gamma_{<} ,
\label{eq:epsstar}
\end{equation}
where $\gamma_{<} = \min \gamma_s \le 1$. In the anisotropic
limit, $\epsilon\rightarrow0$, the TT of the smaller
(larger) FS diverges to negative (positive) values, while their ratio
remains finite and negative,
$(\tau_{s}/\tau_{0,s})/(\tau_{\bar{s}}/\tau_{0,\bar{s}})
\to -\gamma_{\bar{s}} = -1/\gamma_s$.
This can be understood as follows. For
$\epsilon\to0$, the scattering rate $W_a(\theta)$ becomes a $\delta$-function
and, therefore, a particle in the state $s,\mathbf{k}$ can only scatter to the
state $\bar{s},\bar{\mathbf{k}}$, where $\bar{\mathbf{k}}$ is determined by
$\varepsilon_{\bar{s},\bar{\mathbf{k}}}=\varepsilon_{s,\mathbf{k}}$ and
$\theta_{\mathbf{k},\bar{\mathbf{k}}}=0$, see Fig.\ \ref{fig:formulation}(a).
Thus the system decouples into pairs of states, $s,\mathbf{k}$ and
$\bar{s},\bar{\mathbf{k}}$, thereby becoming non-ergodic.
The joint particle density of these two states,
$F_{s,\mathbf{k}}\equiv
N_sf_{s,\mathbf{k}}+N_{\bar{s}}f_{\bar{s},\bar{\mathbf{k}}}$,
cannot change due to scattering and for the e-h case considered here
evolves in time according to
\begin{equation}
\frac{dF_{s,\mathbf{k}}}{dt}=eN_s\mathbf{E}\cdot\mathbf{v}_{s,\bf k}
  [-f_0'(\varepsilon_{s,\bf k})]
  \Big(1-\frac{1}{\gamma_s}\Big).
\label{eq:jpd}
\end{equation}
A steady state is established only for the special case of perfectly
matched FSs, $\gamma_s=1$, leading to finite TTs. Equation
(\ref{eq:trlifetime}) shows that they approach the limit $\tau_s=\tau_{0,s}/2$.
For $\gamma_s\ne1$, however, the joint particle density $F_{s,\mathbf{k}}$
accelerates freely so that the TTs diverge. While this leads to
diverging individual occupations $f_{s,\mathbf{k}}$, the difference
$f_{s,\mathbf{k}}-f_{\bar{s},\bar{\mathbf{k}}}=g_{s,\mathbf{k}}-g_{\bar{s},\bar{
\mathbf{k}}}$
approaches the finite value
 $\tau_{0,s}e\mathbf{E}\cdot\mathbf{v}_{s,\bf k}
[-f_0'(\varepsilon_{s,\bf k})]$, which can be obtained from Eq.\ (\ref{eq:BE})
by considering $\delta$-scattering.

For small $\epsilon>0$, the scattering between $s,\mathbf{k}$ and
$\bar{s},\bar{\mathbf{k}}$ still dominates and scattering to other states can
be treated as a weak perturbation. For the uncompensated case
$\gamma_s\neq 1$, this additional scattering leads to weak
relaxation of $F_{s,\mathbf{k}}$ and thus stabilizes a steady state.
Still, the difference $f_{s,\mathbf{k}}-f_{\bar{s},\bar{\mathbf{k}}}$
relaxes much more rapidly than the individual occupations, so
that $f_{s,\mathbf{k}} \approx f_{\bar{s},\bar{\mathbf{k}}}$ in the steady
state.
The occupation numbers on the same side of the two FSs
are both either enhanced or reduced in comparison to the equilibrium state.
In the e-h case, the electrons at these points have
opposite velocities so that the electrons on one of the FS have to drift
in the ``wrong'' direction. As Fig.\ \ref{fig:tau} shows,
the direction of the drift and thus the signs of the TTs are
set by the majority carriers.

On the other hand, for weak anisotropy, $\epsilon\approx 1$, Fig.\ \ref{fig:tau}
shows that $\tau_s/\tau_{0,s}$ decreases with decreasing
$\epsilon$ regardless of the FS sizes.
Increasing anisotropy favors small-$\theta$ scattering. For the e-h case, this
enhances backscattering since the velocities $\mathbf{v}_{s,\mathbf{k}}$ and
$\mathbf{v}_{\bar{s},\bar{\mathbf{k}}}$ are opposite
and is therefore more efficient in relaxing the current. The enhanced
backscattering is effective for all $\gamma_s$, including the compensated case
of $\gamma_s=1$. At this special value, it is not balanced by the previously
discussed mechanism based on anisotropic scattering so that the TTs remain
finite down to $\epsilon=0$.

In the e-e/h-h case, the carriers from both FSs always drift in the same
direction, as expected. The TTs increase monotonically with 
$\epsilon$, and diverge in the extreme anisotropic limit
$\epsilon\rightarrow2$. The increasing anisotropy favors small-$\theta$
scattering, which for the e-e/h-h case corresponds to forward scattering, and
is thus increasingly inefficient at relaxing the current.

We finally turn to the consequences of additional isotropic
intraband scattering. If we include $W_i$ in Eq.\ (2), the TTs become
\begin{equation}
\tau_{s} = \tau_{0,s}\,
  \frac{1-\frac{1}{\gamma_{s}}\,\frac{1-\epsilon}{1+x_{\bar{s}}}}
  {1-\frac{1-\epsilon}{1+x_{s}}\,\frac{1-\epsilon}{1+x_{\bar{s}}}} ,
\end{equation}
where the bare lifetimes now consist of intra- and interband contributions,
\begin{equation}
\frac{1}{\tau_{0,s}} = \frac{1}{\tau_{0,s}^{(a)}}
  + \frac{1}{\tau_{0,s}^{(i)}}
  \equiv N_{\bar{s}} \left\langle W_{a}(\theta)\right\rangle_{\theta}
  + N_s\, W_i ,
\end{equation}
and $x_{s}\equiv\tau_{0,s}^{(a)}/\tau_{0,s}^{(i)}$ is the ratio of the
lifetimes due to inter- and intraband scattering.
If we assume equal densities of states at the two FSs for simplicity we
recover the previous expression (\ref{eq:trlifetime}) for the TTs
with a renormalized anisotropy parameter
\begin{equation}
\epsilon \rightarrow \frac{\epsilon+x}{1+x} ,
\end{equation}
where $x = W_{i}/\langle W_{a}(\theta)\rangle_{\theta}$. Thus in this
case the only effect of isotropic intraband scattering is to reduce the
range of the anisotropy parameter to $x/(1+x)\le\epsilon\le2-x/(1+x)$.
Note that negative TTs still occur provided that $x<1/\gamma_{<}-1$.
Since the inclusion of isotropic intraband scattering essentially leads 
to a renormalization of the anisotropy parameter,
we ignore intraband scattering in the following discussion
of the transport coefficients.

\section{Transport coefficients}
\subsection{Resistivity}

\begin{figure}[tb]
\includegraphics[width=\columnwidth]{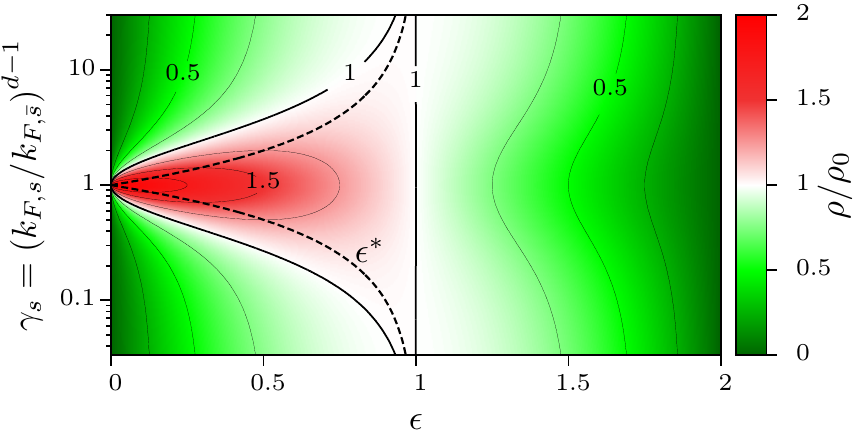}
\caption{(Color online) Resistivity in terms of its isotropic limit,
$\rho_{0}\equiv\rho|_{\epsilon=1}$,
as a function of $\epsilon$ and
$\gamma_{s}=\left(k_{F,s}/k_{F,\bar{s}}\right)^{d-1}$.
At $\epsilon=\epsilon^*$ (dashed curve), the TT of the minority
carriers changes sign and the resistivity has a maximum as a function of
$\epsilon$.}
\label{fig:res}
\end{figure}

The resistivity can be obtained
from the TTs $\tau_s$.\cite{Ziman1972} 
We present the resistivity relative to its isotropic limit,
$\rho_{0}\equiv\rho|_{\epsilon=1}$, which coincides with
the result one would obtain by approximating the TTs by
bare lifetimes. We obtain
\begin{eqnarray}
\frac{\rho}{\rho_{0}}
& = & \frac{v_{F,1} k_{F,1}^{d-1} \tau_{0,1} + v_{F,2} k_{F,2}^{d-1} \tau_{0,2}}
    {v_{F,1} k_{F,1}^{d-1} \tau_1 + v_{F,2} k_{F,2}^{d-1} \tau_2} \nonumber \\
& = & \frac{2-\epsilon}
  {1+\frac{1-\epsilon}{\epsilon}\left(1-\frac{2}{\gamma_1+\gamma_2}\right)} .
\label{eq:res}
\end{eqnarray}
The ratio $\rho/\rho_0$ is plotted in Fig.~\ref{fig:res}. The
figure shows that the anisotropy has a large effect on the
resistivity, especially in the e-h case.
Although minority carriers give a negative contribution to the current
for $\epsilon<\epsilon^{*}$,
the total current in the direction of $\mathbf{E}$
is always positive.
In the uncompensated case ($\gamma_s\ne1$) the competition between
the two anisotropy effects, the usual enhancement of the resistivity due
to backscattering and the reduction due to anisotropic scattering 
causes a maximum of $\rho/\rho_0$ as a function of $\epsilon$
at $\epsilon=\epsilon^*$.

Consistent with previous investigations\cite{Fanfarillo2012} we find that in 
compensated e-h systems $\rho/\rho_0$ exhibits an enhancement up to a factor
of two due to the usual backscattering.
In uncompensated e-h systems, however, anisotropy of the scattering causes
a strong reduction of the resistivity below $\epsilon=\epsilon^*$, which 
occurs already at weak anisotropy, if the mismatch between the FS radii is large.

\bigskip\subsection{Hall coefficient}

\begin{figure}[tb]
\includegraphics[width=\columnwidth]{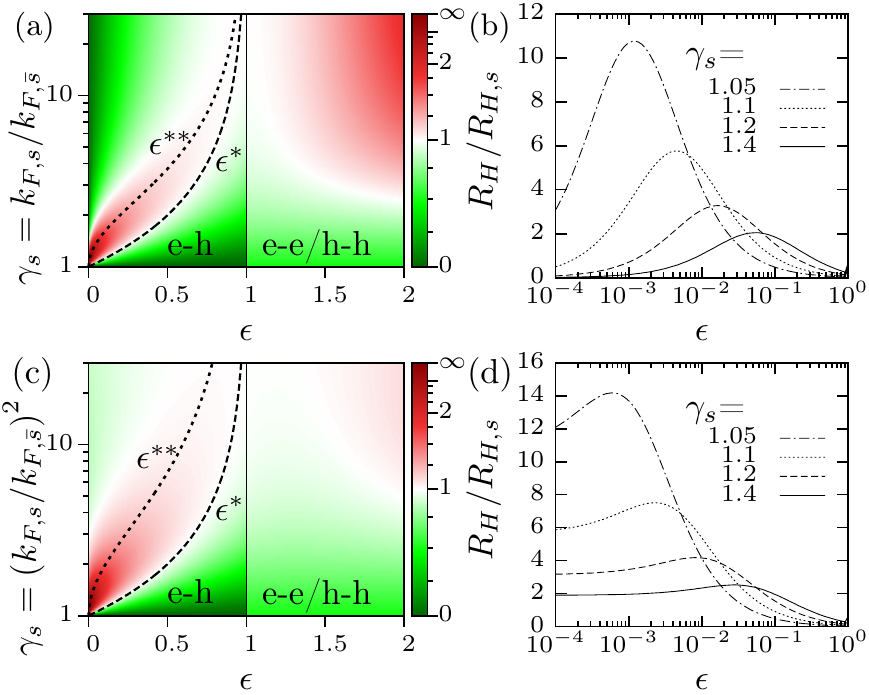}
\caption{(Color online) (a) Hall coefficient $R_H$ for a 2D system
as a function of $\epsilon$ and $\gamma_{s}$, in units of the Hall coefficient
$R_{H,s}$ of FS $s$. (b) The same quantity as a function of $\epsilon$ on a
logarithmic scale for various $\gamma_s$.
(c), (d) The same as in panels (a), (b) for a 3D system.
We only plot $R_H$ in terms of $R_{H,s}$ for the larger FS ($\gamma_s\ge 1$);
the result in terms of the smaller FS is easily obtained from Eq.\
(\ref{eq:hall}). Note that the limit $\epsilon\to 1$ in Eq.\ (\ref{eq:hall})
is different for the e-h and e-e/h-h cases.
The characteristic anisotropy level $\epsilon^{*}$ where the TT of
the minority carriers changes sign is indicated by a dashed line.
The dotted line marked by $\epsilon^{**}$ indicates the maximum of
$R_H/R_{H,s}$ as a function of $\epsilon$.}
\label{fig:hall}
\end{figure}

The Hall coefficient of a two-band system with
isotropic dispersion obeys \cite{Ziman1972}
\begin{equation}
R_{H} = \pm \frac{1}{ec}\, \frac{n_s\mu_s^2  + \eta_{s\bar{s}}\,
n_{\bar{s}}\mu_{\bar{s}}^2}{(n_s\mu_s+ n_{\bar{s}}\mu_{\bar{s}})^2} ,
\label{eq:hall_general}
\end{equation}
where $\mu_s=e\tau_s v_{F,s}/k_{F,s}$ and $n_s\propto k_{F,s}^d$
are the mobility and particle number of FS $s$, respectively, and
the upper (lower) overall sign pertains to a holelike (electronlike) FS $s$.
With Eq.\ (\ref{eq:trlifetime}) we obtain
\begin{equation}
\frac{R_{H}}{R_{H,s}} =
  \frac{1 + \eta_{12}\, \gamma_{s}^{\frac{4-3d}{d-1}}
  \left[\frac{1-\gamma_{s}(1-\epsilon)}
  {1-\gamma_{s}^{-1}(1-\epsilon)}\right]^{2}}
  {\left[1+\gamma_{s}^{-2}
  \frac{1-\gamma_{s}(1-\epsilon)}{1-\gamma_{s}^{-1}
  (1-\epsilon)}\right]^{2}} ,
\label{eq:hall}
\end{equation}
where $R_{H,s} \propto 1/n_s$ is the Hall coefficient of FS $s$.
In the e-h case, the electrons and holes contribute with different signs to the
Hall coefficient, irrespective of the signs of the TTs. 
The ratio $R_H/R_{H,s}$ is plotted for the 2D and 3D cases in Fig.\
\ref{fig:hall}. The figure shows that $R_H/R_{H,s}$ is strongly affected by the
anisotropy, implying that approximating the TT by the lifetime\cite{qplifetime}
is not sufficient. In particular, we find a maximum at
\begin{equation}
\epsilon^{**} \equiv
  \frac{(\gamma_{s}-1)\Big(1-\gamma_{s}^{-\frac{1}{d-1}}\Big)}
  {\gamma_{s}^{-\frac{1}{d-1}}+\gamma_{s}} ,
 \label{eq:epsstarstar}
\end{equation}
which corresponds to equal magnitude but opposite sign of the electron and
hole mobilities, $\mu_e=-\mu_h$.
At the maximum the Hall coefficient assumes the value
$R_{H}/R_{H,s} = 1/[1-\gamma_{s}^{-d/(d-1)}]$,
which diverges for $\gamma_{s}\to1$ so that for nearly compensated
e-h systems anisotropic scattering can cause a huge enhancement of the Hall
effect in agreement with Fanfarillo \textit{et al.}\cite{Fanfarillo2012}
Going beyond Ref.\ \onlinecite{Fanfarillo2012}, we predict that
an extremum in the Hall coefficient should be observed when the anisotropy
parameter $\epsilon$ is tuned through the value $\epsilon^{**}$.
According to Eq.\ (\ref{eq:epsstarstar}) and Fig.\ \ref{fig:hall}, the
anisotropy required to reach the extremum decreases ($\epsilon^{**}$
increases) for a larger mismatch between the FS radii.
We speculate that this is the reason why among the pnictides $\mathrm{LiFeAs}$
and $\mathrm{LiFeP}$, which have rather poor nesting, show the most
pronounced extremum in the Hall coefficient as a function of
temperature.\cite{key-16,key-17}
It is remarkable that also in 122 pnictides an extremum in the Hall coefficient
is observed for sufficiently strongly doped systems.\cite{122Hallextremum}

\bigskip\subsection{Magnetoresistance}

\begin{figure}[t]
\includegraphics[width=\columnwidth]{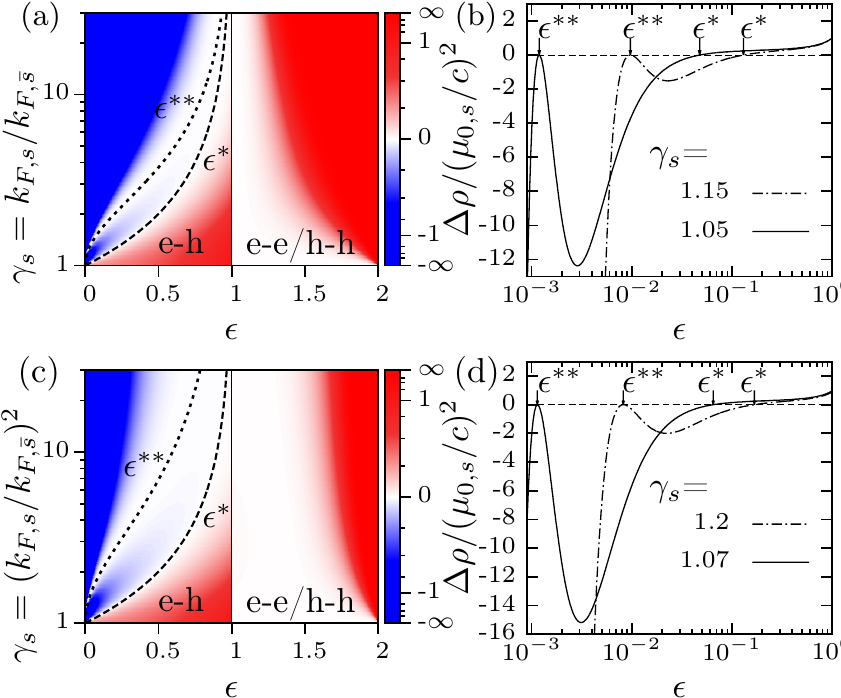}
\caption{(Color online) (a) Magnetoresistance coefficient for the 2D
system in units of
$(\mu_{0,s}/c)^2$ as a function of $\epsilon$ and $\gamma_{s}$ and (b) as a
function
of $\epsilon$ for two values of $\gamma_s$.
(c), (d)  The same as in panels (a), (b) for a 3D system.
At $\epsilon=\epsilon^{*}$ (dashed curve) and $\epsilon^{**}$ (dotted curve) the
magnetoresistance is zero.}
\label{fig:MR}
\end{figure}

\begin{figure}[t]
\includegraphics[width=\columnwidth]{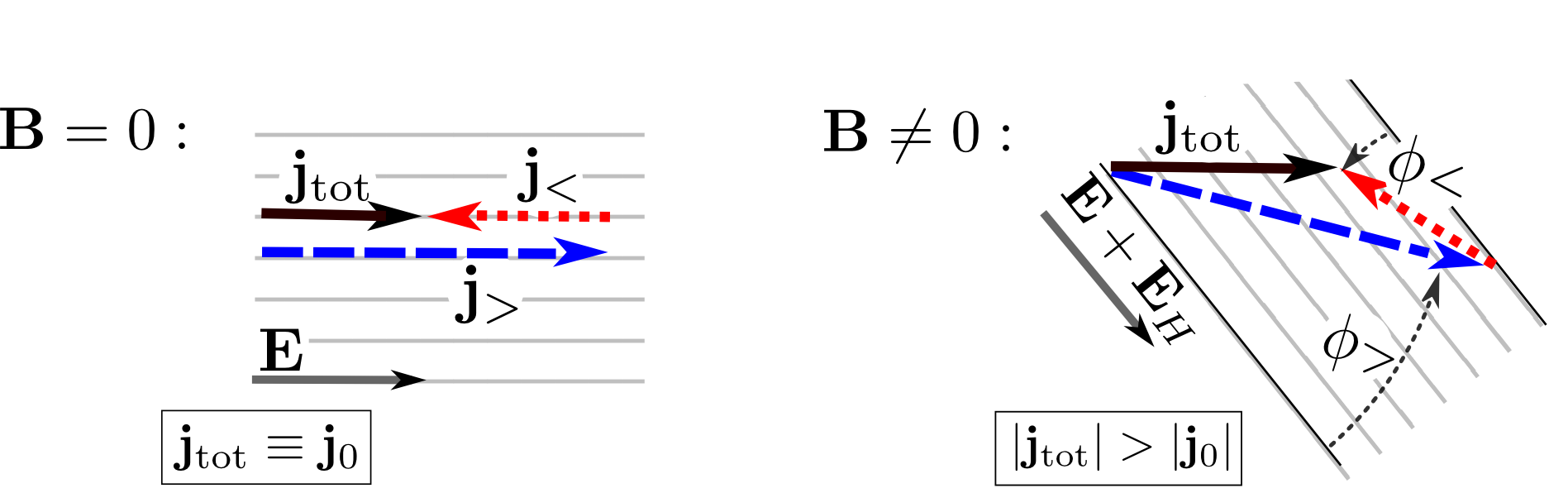}
\caption{(Color online) Illustration of the negative magnetoresistance below
$\epsilon^{*}$. The magnetic field $\mathbf{B}$ 
rotates the current contribution of the smaller (larger)
FS, $\mathbf{j}_<$ ($\mathbf{j}_>$), by the angle $\phi_<$ ($\phi_>$), which
leads to an \emph{increase} of the total current
$\mathbf{j}_{\mathrm{tot}}=\mathbf{j}_>+\mathbf{j}_<$ if $\phi_<\ne\phi_>$.
The induced Hall electric field $\mathbf{E}_H$ has no effect on the
magnetoresistance but ensures that the total current points in the
direction parallel to the applied electric field $\mathbf{E}$.}
\label{fig:MRexpl}
\end{figure}

The magnetoresistance coefficient
\begin{equation}
\Delta\rho \equiv \frac{\rho(B)-\rho(0)}{\rho(0) B^{2}}
\end{equation}
is obtained from the standard expression \cite{Ziman1972} 
\begin{equation}
\Delta\rho = \frac{n_1\mu_1n_2\mu_2}{(n_1\mu_1+n_2\mu_2)^2}\,
  \bigg(\frac{\mu_s}{c}
  -\eta_{s\bar{s}}\frac{\mu_{\bar{s}}}{c}\bigg)^2 ,
\end{equation}
to leading order in the magnetic field $\mathbf{B}$. In terms of the TTs we
find
\begin{equation}
\frac{\Delta\rho}{(\mu_{0,s}/c)^2} =
  \frac{\frac{\tau_1}{\tau_{0,1}}\frac{\tau_2}{\tau_{0,2}}}
  {\big(\frac{\tau_1}{\tau_{0,1}}\gamma_1
  + \frac{\tau_2}{\tau_{0,2}}\gamma_2\big)^{\!2}}\,
  \bigg(\frac{\tau_s}{\tau_{0,s}}
  - \eta_{12}\,\gamma_s^{\frac{2-d}{d-1}}\,
  \frac{\tau_{\bar{s}}}{\tau_{0,\bar{s}}} \bigg)^{\!\!\!2} ,
\end{equation}
where $\mu_{0,s}=e\tau_{0,s} v_{F,s}/k_{F,s}$ is the bare mobility for FS $s$.
Results are plotted in Fig.\ \ref{fig:MR}. The magnetoresistance is negative
for $\epsilon<\epsilon^{*}$, where one TT becomes negative. Figure
\ref{fig:MRexpl} illustrates the
connection between negative TT and negative magnetoresistance.
If the TT of the minority carriers is negative (positive), the
current contributions $\mathbf{j}_<$ and $\mathbf{j}_>$ of minority and
majority carriers, respectively, point in opposite directions (the same
direction) for $\mathbf{B}=0$. If a magnetic field is applied, the contributions
$\mathbf{j}_<$ and $\mathbf{j}_>$ are rotated due to the Lorentz force, under
the constraint that the total current in the transverse direction vanishes.
However, as long as $\mathbf{j}_<$ and $\mathbf{j}_>$ are not rotated by the
same angle, they are no longer parallel and their vector sum is thus
\emph{larger} (\emph{smaller}) in
absolute value than for $\mathbf{B}=0$. Hence, the magnetoresistance is
negative (positive). A special point is $\epsilon=\epsilon^{**}$, where the two
angles are equal, the total current is unchanged, and the
magnetoresistance vanishes.

Since many materials of interest, such as the iron pnictides, have
more than two FSs, we briefly discuss this case.
In the presence of additional isotropic FSs, the magnetoresistance
coefficient is determined by a sum over all pairs of FSs,
\begin{equation}
\Delta\rho= \frac{\frac{1}{2}\sum_{i,j=1} n_i\mu_i n_j\mu_j\,
  \big(\frac{\mu_i}{c}-\eta_{ij}\frac{\mu_{j}}{c}\big)^2}
   {\big(\sum_{i=1} n_i\mu_i\big)^2}.
\end{equation}
Each pair of FSs with opposite signs of the TTs gives a
negative contribution to the magnetoresistance. Hence, when more than two
FSs are present, the appearance of negative TTs does not
necessarily lead to a negative magnetoresistance coefficient, although it can
be significantly reduced. This could be relevant for the observation of
an unexpectedly small magnetoresistance in certain
pnictides.\cite{Kim}

\bigskip\section{Conclusions}

Anisotropic single-particle scattering
between electron and hole FSs causes the transport coefficients
to differ dramatically from the expectation based on bare lifetimes.
The unexpected behavior is especially pronounced in the regime where
the minority carriers have negative TTs. Here the 
magnetoresistance is negative, the Hall coefficient exhibits an extremum,
and the resistivity decreases upon increasing the anisotropy.
The degree of anisotropy required for this to occur decreases for
increasing ratio between the FS areas.
This effect does not depend on a particular microscopic origin for the
anisotropic scattering~\cite{Fanfarillo2012} and is distinct from
carrier drag due to the two-particle electron-hole interaction.\cite{key-2} 

Some general conclusions may be drawn.
Close to perfect nesting, negative TTs are restricted to the limit of
extreme scattering anisotropy and thus should
only become evident in the transport just above the excitonic
instability. Significant doping may therefore be required to observe the most
striking effects. It is, however, encouraging that in the pnictides 
the magnetoresistance is rather small,\cite{Kim} while the Hall
coefficient is strongly enhanced close to the spin-density-wave
transition,\cite{Fang2009,Shen2011} 
consistent with our predictions. In contrast to what is stated in 
Ref.\ \onlinecite{Fanfarillo2012}, we show that the Hall coefficient enhancement 
can be explained within the semiclassical approach. Moreover
we predict that the Hall coefficient exhibits an extremum when
the system is tuned through a characteristic degree of anisotropy, which
becomes weaker for larger mismatch between the FS radii. This could
explain the appearance of
an extremum in the Hall coefficient as a function of temperature in
strongly doped 122 pnictides,\cite{Fang2009} as well as in $\mathrm{LiFeAs}$
and $\mathrm{LiFeP}$,\cite{key-16,key-17} which show rather poor
nesting. However, the most decisive test of negative TTs resulting from
anisotropic interband scattering would be the measurement of a
negative magnetoresistance.

\acknowledgements

Financial support by the Deutsche Forschungsgemeinschaft through Research
Training Group GRK 1621 is gratefully acknowledged. We also thank Dirk
Morr for a useful discussion.\\

\end{document}